\documentclass[twocolumn,secnumarabic,amssymb, nobibnotes, aps, prd]{revtex4}
\usepackage{graphicx}

\begin{document}
\title{A possibility of persistent voltage observation in a system of asymmetric superconducting rings}
\author{A. A. Burlakov, V. L. Gurtovoi, A. I. Ilin, A. V. Nikulov, V. A. Tulin}
\affiliation{Institute of Microelectronics Technology and High Purity Materials, Russian Academy of Sciences, 142432 Chernogolovka, Moscow District, RUSSIA} 

\begin{abstract} A possibility to observe the persistent voltage in a superconducting ring of different widths of the arms is experimentally investigated. It was earlier found that switching of the arms between superconducting and normal states by ac current induces the dc voltage oscillation in magnetic field with a period corresponding to the flux quantum inside the ring. We use systems with a large number of asymmetric rings connected in series in order to investigate the possibility to observe this quantum phenomenon near the superconducting transition where thermal fluctuations switch ring segments without external influence and the persistent current is much smaller than in the superconducting state.
 \end{abstract}

\maketitle

\narrowtext

\section{Introduction}
It is well known that an electrical current $ I_{0}$ circulating in a ring with a non-zero resistance $R_{l} > 0$ is maintained $R _{l} I_{0} = -d\Phi /dt$ by the Fadaday electric field $E = - dA/dt = -l^{-1} d\Phi /dt$. The current would rapidly decay $I(t) = I_{0}\exp -t/\tau_{RL}$ during the relaxation time $\tau_{RL} = L/R_{l} $ when a magnetic flux inside the ring $\Phi $ does not change with time $ d\Phi /dt = 0$. $l = 2\pi r$ is the circumference of the ring; $L$ is the inductance of the ring. However, the measurements \cite{PCScien09, PCPRL09,PCScien07} demonstrate that an electrical current cannot decay in spite of $R _{l} > 0$ and $ d\Phi /dt = 0$. These mesoscopic quantum phenomena, known as persistent current, are observed in a normal metal ring \cite{PCScien09, PCPRL09} at temperatures enough low and at temperatures corresponding to the fluctuation region of the superconducting transition \cite{PCScien07}. 

The quantum theories \cite{KulikS,KulikN,Riedel88,Riedel92,Riedel93} describe these phenomena \cite{PCScien09,PCPRL09,PCScien07} as a consequence of the momentum quantization 
$$\oint _{l}dl p  = \oint _{l}dl (mv + qA) = \oint_{l}dl \hbar \nabla \varphi = n2\pi \hbar \eqno{(1)}$$ 
of quantum particles, i.e. particles described by the wave function $\Psi  = |\Psi| \exp i\varphi $, and the Aharonov-Bohm effect, connected with the dependence of the canonical momentum $p = mv + qA$ of a charged $q$ particle on the magnetic vector potential $A$. The direction and value of the persistent current corresponding to thermodynamic equilibrium change periodically with a magnetic flux $\Phi = \oint _{l}dl A = BS$ inside the ring of an area $S = \pi r^{2}$ \cite{PCScien09,PCPRL09,PCScien07}. The $I _{p}(\Phi /\Phi _{0})$ oscillation period  corresponds to the flux quantum $\Phi _{0} = 2\pi \hbar /q$ \cite{PCScien09,PCPRL09,PCScien07}. 

The persistent current of electrons (which are fermions with the charge $q = e$) observed in a normal metal ring \cite{PCScien09,PCPRL09} oscillates with the period $\Phi _{0} = 2\pi \hbar /e \approx 41.4 \ Oe \ \mu m^{2}$ and its amplitude cannot exceed $M^{1/2}I _{0}$ \cite{Riedel93}, where $M$ is the number of one-dimensional channels \cite{Riedel93} and $I _{0}$ is the current of each channel. It can be said that the current $I _{0}$ at $T = 0$ is the current of single electron $I _{0} \approx  ev _{F}/2\pi r $ on the Fermi level $n _{F} = \pm mv _{F}r/\hbar $ \cite{Riedel93} because the oppositely directed currents of electrons occupying other permitted levels (1) compensate for each other \cite{Riedel88}. The amplitude of the $I _{p}(\Phi /\Phi _{0})$ oscillations observed in measurements of real normal metal rings \cite{PCScien09} is appreciably lower than the current of a single electron ($ ev _{F}/2\pi r \approx 50 \ nA$ at typical radius $r \approx 0.5 \ \mu m$ and the Fermi velocity $v _{F} \approx  10^{6} \ m/c $) because of electron scattering and non-zero temperature of measurements \cite{Riedel93}. The amplitude of the oscillations $I _{p}(\Phi /\Phi _{0})$  with the period $\Phi _{0} = 2\pi \hbar /2e \approx 20.7 \ Oe \ \mu m^{2}$ observed in measurements of a superconducting ring is appreciably higher $I _{p} = 2eN _{s}v _{n}/2\pi r = 2esn _{s}v _{n}$ even near the critical temperature $T \approx T _{c}$  \cite{PCScien07} because all $N _{s} = Vn _{s}$ pairs (bosons with the charge $q = 2e$) in the ring of a macroscopic volume $V = s 2\pi r$ have the same velocity $v _{n}$ corresponding to the minimum energy $ \propto v _{n}^{2} \propto (n - \Phi /\Phi _{0})^{2}$. The quantum theories \cite{Riedel92,Riedel93} describe the quantum oscillations $I _{p}(\Phi /\Phi _{0})$ and temperature dependence of their amplitudes both for normal metal \cite{PCScien09} and superconducting \cite{PCScien07} rings. But these theories do not answer two important questions: "Why does the persistent current not decay in spite of non-zero resistance?" and "Could the persistent current induce a persistent voltage in the ring arms of different resistance?"   

Two possible answers to the first question are considered in the paper \cite{Kulik75}. According to the first answer, assumed by the authors of \cite{PCScien09} and \cite{Birge09}, the persistent current can flow without dissipating energy in spite of non-zero electric resistance. Contrary to this preposterous assumption \cite{Birge09} the author \cite{KulikN} emphasized, as far back as 1970, that weak energy dissipation should not result in the disappearance of the persistent current because this equilibrium phenomenon corresponds to the minimum free energy. According to this second answer the persistent current observed at non-zero resistance is the Brownian motion \cite{FeynmanLec} similar to the Nyquist \cite{Nyquist} (or Johnson \cite{Johnson}) equilibrium electric current. It is argued in \cite{Kulik75} that this second answer rather than the first one agrees with experimental data. 

The antithetical answers to the first question also presume opposing answers to the second one. The authors \cite{PCScien09} use the analogy to stationary electron orbits of an atom as a sole argument for their interpretation of the persistent current as dissipationless. But the rings used for the observation of the persistent current  \cite{PCScien09,PCPRL09,PCScien07} basically differ from atoms. Rings have an electric resistance $R _{l}$ and it is possible to make a ring with different resistance of its arms $ R _{n} > R _{w}$,   $R _{n} + R _{w} = R _{l}$. It is well known that the potential difference $V_{n} = -V_{w} = 0.5(R _{n} - R _{w})I$ is observed in both arms of the ring $l_{n}$, $l_{w}$ when a conventional electric current $(R _{n} + R _{w})I = -d\Phi /dt$ maintained by the Fadaday electric field $E = - dA/dt$ circulates along the ring circumference $l = l_{n} + l_{w}$. The sign of the voltage $V_{w}$ on the arm with smaller resistance $ R _{w}$ is opposite to the $I$ sign. But the current  is observed in accordance with the Ohm's law $I = s_{n}j_{n} = s_{n}\rho_{n} E_{n} = s_{w}j_{w} = s_{w}\rho_{w} E_{w}$ and its direction corresponds to the direction of an electric field $E = -\nabla V - dA/dt$ in both arms, $E_{n} = - l_{n}^{-1}0.5(R _{n} - R _{w})I - l^{-1}d\Phi /dt$ and $E_{w} = l_{w}^{-1}0.5(R _{n} - R _{w})I - l^{-1}d\Phi /dt$, due to the non-zero Fadaday voltage $-d\Phi /dt$. The persistent current, in contrast to the conventional circular current, is observed at $d\Phi /dt = 0$ and it should be directed opposite to the electric field $E = -\nabla V$ in one of the arms at $V_{n} = -V_{w} \neq 0$. Therefore it may seem that no persistent voltage can arise, basing on the idea that no direct electric current can flow against the direct electric field $E = -\nabla V$ in the ring arm. 

But the experimental data \cite{Kulik75,Letter07} have disproved this. The persistent current flowing against the electric field $E = -\nabla V$ is observed in measurements of the Little-Parks oscillations \cite{Letter07}. These oscillations of the resistance observed by W. A. Little and R. D. Parks as far back as 1962 \cite{LP1962} is the first experimental evidence of $I _{p} \neq 0$ at  $R _{l} > 0$. Another phenomenon disproving the idea that the persistent current cannot flow opposite to the electric field was discovered forty years later \cite{Dubonos02}. The observations \cite{Kulik75,Letter07,Dubonos02} of the dc voltage $V _{dc}$ on the arms of asymmetric rings, the sign and value of which oscillate in magnetic field like the persistent current $\overline{I _{p}}$, give unambiguous evidence that the averaged electric current $\overline{I _{p}}$ flows against the dc electric field $E = -\nabla V _{dc}$ in one of the arms. In terms of quantum mechanics these paradoxical phenomena can be described as a consequence of switching between superconducting states with different connectivity of the wave function \cite{JLTP1998,PRB2001}.       

In order to verify the possibility of persistent voltage existence the amplitude of oscillations $V _{p}(BS/\Phi _{0}) \propto I _{p}(BS/\Phi _{0})$ should be increased to a measurable value. In one of the first attempts to observe the persistent current in normal metal rings the authors \cite{Levy1990} measured a magnetization response of an array of 10 million copper rings. Their strategy was to add together many small signals to create a much larger, measurable signal. This attempt was not quite successful because the persistent current in each normal metal ring has a random sign and therefore the total signal was proportional only to the square root $N^{1/2}$ of the number of rings $N$. In contrast to this case the persistent current in each superconducting ring of identical radius has the same sign at any value of magnetic field $B$. Therefore it can be expected that the persistent voltage of a system of asymmetric superconducting rings connected in series can increase in proportion to the number of rings $N$. The proportionality $V _{p,N}(BS/\Phi _{0}) \approx  NV _{p,1}(BS/\Phi _{0})$ implies that the voltage $V _{p,N}(BS/\Phi _{0})$ can be made measurable even at very small persistent voltage $V _{p,1}(BS/\Phi _{0})$ of a single ring. For example, in a system of $N = 10^{7}$ rings the oscillations $V _{p,N}(BS/\Phi _{0})$ with an amplitude exceeding the measurable voltage $V _{mab} \approx 20 \ nV = 2 \ 10^{-8} \ V$ can be observed even at the extremely small persistent voltage value $V _{p,1} \approx 10^{-14} \ V$.

Our experimental facilities has allowed the fabrication of systems with a much smaller number of asymmetric aluminium rings, $N = 110$  and  $N = 1080$, Fig.1. The measurements of these systems showed that the number of rings $N = 1080$ is not sufficient to observe the persistent voltage above the superconducting transition. Nevertheless we could observe oscillations of the dc voltage at temperatures corresponding to the lower part of this transition. The results of our measurements corroborate the proportionality $V _{p,N}(BS/\Phi _{0}) \propto N$ and allow the estimation of the number $N$ required for the observation of persistent voltage above the superconducting transition.

\section{Experimental Details}  
The structures consisting of 110 and 1080 series-connected asymmetric rings were fabricated by the lift-off method by depositing a thin aluminium film $d \approx 20 \ nm$ thick onto a Si substrate. The lithography was performed using a JEOL-840A scanning electron microscope transformed into a laboratory electron lithograph by the NANOMAKER program package. All rings are of the same inner radius $r \approx  0.9 \ \mu m$ and semi-ring widths $w_{w} \approx  0.4 \ \mu m$, $w_{n} \approx  0.2 \ \mu m$, Fig.1. The distance between the rings and the width of the strips connecting them is $\approx 0.6 \ \mu m$. The cross sections of the semi-rings are $s_{w} = w_{w}d \approx  0.008 \ \mu m^{2}$ and $s_{n} = w_{n}d \approx  0.004 \ \mu m^{2}$. With the London penetration depth and the correlation length in aluminium film being $\lambda _{L}(T) \approx  0.05 \ \mu m (1 - T/T_{c})^{-1/2}$ and $\xi (T) = \xi (0)(1 - T/T_{c})^{-1/2} \approx  0.13 \ \mu m (1 - T/T_{c})^{-1/2}$ \cite{Letter07}, the cross sections correspond to a weak screening $s_{n} < s_{w} < \lambda _{L}^{2}(T)$ and the structure is one-dimensional $d, w _{n}, w _{w} < \xi (T) $ at the temperature of measurements $T > 0.9T_{c}$. The midpoint of the resistive transition corresponds to $T_{c} \approx  1.36 \ K$ for a 110 ring system and $T_{c} \approx  1.37 \ K$ for a 1080 ring system; the width of the transition is $\Delta T_{c}(0.1 \div 0.9R_{n}) \approx  0.02 \ K$ for the booth systems, and the maximum slope is $dR/d(T-T_{c}) \approx  30000 \ \Omega /K$ and $dR/d(T-T_{c}) \approx  280000 \ \Omega /K$. The resistance of the 110 and 1080 rings structures in the normal state is $R_{n} \approx  970 \ \Omega $ (the resistance of each ring is $R_{n,1} \approx  8.7 \ \Omega $ )  and $R_{n} \approx  8000 \ \Omega $ (the resistance of each ring is $R_{n,1} \approx  7.4 \ \Omega $ ). The resistance per square is $ \rho _{n}/d \approx  1.4 \ \Omega /\diamondsuit $, $\rho _{n} \approx  2.8 \ 10^{-8} \ \Omega m$, and the resistance ratio is $R(300 \ K)/R(4.2 \ K) \approx  1.7$.

\begin{figure}[b]
\includegraphics{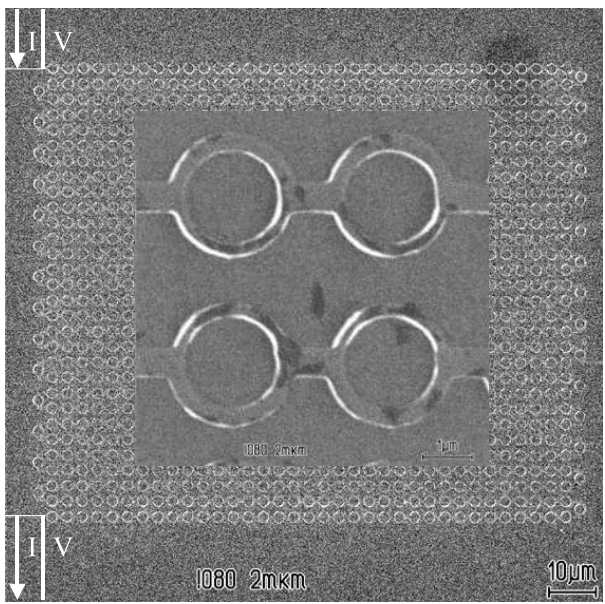}
\caption{\label{fig:epsart} System of 1080 series-connected asymmetric aluminum rings with the same inner radius $r \approx  0.9 \ \mu m$ and the semi-ring widths $w_{w} \approx  0.4 \ \mu m$, $w_{n} \approx  0.2 \ \mu m$. The voltage was measured between the V-V contact pads, and for the current the I-I contact pads were used. The inset at the centre shows a system fragment of a large scale.}
\end{figure}

The measurements were performed in a glass cryostat using $He^{4}$ as a cooling agent. By pumping, the temperature could be reduced to $1.2 \ K$. The voltage was measured across the potential contact pads $V-V$ (Fig.1) by an instrumental amplifier with a gain of 1000 and an input-normalised noise level of $20 \ nV$ in the frequency band from 0 to 1 Hz. Then, the signal was supplied to an SR560 preamplifier (Stanford Research Systems), which was used for additional amplification and formation of the required signal frequency band by low-pass and high-pass filters. Because of the galvanic connection between the measured sample located at low temperature $T \approx 1 \ K$ and the measuring system located at room temperature $T \approx 300 \ K$, a thermal electric noise disturbs the thermodynamic equilibrium of ring system. In this work, we used a filtration system consisting of low-temperature $\pi $-filters and coaxial resistive twisted pairs with attenuation at high frequencies to reduce this noise and to thermalize samples at low temperatures. Owing to the filtration, the frequency of the transmission band of each conductor leading to the sample was equal to $30 \ kHz$, and the suppression 70 to 80 dB was achieved at frequencies in the range $100 \ MHz$ to $10 \ GHz$.   

The temperature was measured by a calibrated resistance thermometer ($R(300 K) = 1.5 \ k\Omega $) with a measuring current of $0.1 \ \mu A$. An external current $I_{ext}$, constant, sine or noise, was supplied to the current contact pads $I - I$ (Fig.1). The constant $I_{ext}$ current from $0.1 \ nA$ to $2 \ \mu A$ was used to measure the resistance versus magnetic field $R(B)$ (the Little-Parks oscillations) and temperature $R(T)$. The sine $I_{ext} = I _{sine} = I _{f} \sin (2\pi f )$ and noise $ I_{ext} = I _{noise} = \int _{fmin}^{fmax} df I _{f} \sin (2\pi f + \phi _{f})$ currents were used to calibrate  the ring systems as a noise detector i.e. in order to obtain the dependence of the amplitude of dc voltage oscillations in magnetic field $V_{dc}(B)$ on the amplitude $ I _{f}$ or $(\int _{fmin}^{fmax} df I _{f}^{2})^{1/2} =  \overline{2I _{noise}^{2}}^{1/2}$. The amplified voltage taken from the sample and the signals proportional to the current passing through the sample, to the magnetic field, and to the temperature, were simultaneously digitized by a 16-digit A/D converter with eight differential inputs.

\section{Experimental Results and Discussion}
The persistent voltage, i.e. the dc voltage $V _{p}(BS/\Phi _{0}) \propto I _{p}(BS/\Phi _{0})$ at the thermodynamic equilibrium, may be observed only in the temperature region near $T _{c}$ corresponding to the resistive transition where $0 < R(T) < R _{n}$ because of switching by thermal fluctuations of ring segments between superconducting and normal states. The resistance of homogeneous one-dimensional superconductor changes from zero $R(T) \approx 0$ to the resistance in the normal state $R(T) \approx R _{n}$ in a narrow interval estimated with the standard scaling factor $\varepsilon _{c} = 2.58 Gi _{1D} $ for the temperature variable $\varepsilon = T/T _{c} - 1$, see Fig.19 of \cite{FlucRPP75} and Fig.1 \cite{FlucLL1972}, where $Gi _{1D} = (k _{B}T _{c}/4\pi \mu _{0}H _{c}(0)^{2}s\xi (0))^{2/3} $ is sometimes called the Ginzburg number. A mean free path $l$ of the aluminum systems used in our work is short $l < \xi _{0}$ and therefore their resistive transition should be described by a theoretical dependence obtained in the dirty limit, Fig.1a \cite{FlucLL1972}.  According to this dependence the resistance should change, for example, from $R = 0.1R _{n}$ to $R = 0.9R _{n}$ in the interval $\varepsilon _{0.9} - \varepsilon _{0.1} \approx 3 \times  \varepsilon _{c}$. The Ginzburg number of the one-dimensional aluminum structure used in our work is approximately equal to $Gi _{1D} \approx 0.0008$ at its parameters: $T _{c} \approx 1.36 \ K$, the average cross sections of the semi-rings $(s_{w} + s_{n})/2 \approx  0.006 \ \mu m^{2}$, the thermodynamic critical field $H _{c}(0) \approx 100 \ Oe $ and the correlation length $\xi (0) \approx  0.13 \ \mu m $ \cite{Letter07} extrapolated to the zero temperature. Our measurements showed that the width of resistive transition of a single ring agrees with the expected value $\varepsilon _{0.9} - \varepsilon _{0.1} \approx 3 \times 2.58 Gi _{1D} \approx 0.006$ whereas the resistive transition of ring systems is wider. The latter suggests that these systems are not homogeneous.  

The persistent voltage should exponentially decrease above the superconducting transition $\varepsilon  \gg \varepsilon _{c}$ because the persistent current decrease with temperature increase $I _{p}(T) \propto \exp -(T/T _{c} - 1)^{1/2}$ \cite{KulikS}. Below the transition at $-\varepsilon  \gg \varepsilon _{c}$ the amplitude $I _{p,A} $ of the persistent current oscillations $I _{p}(\Phi /\Phi _{0})$ increases with temperature decrease $I _{p,A} = I _{p,A}(0)(1 - T/T _{c}) = I _{p,A}(0) (-\varepsilon )$ \cite{PCJETP07}. The persistent voltage could be expected to exponentially decrease with going away from $T _{c}$ also at $-\varepsilon  \gg \varepsilon _{c}$ because of an exponential decrease of the resistance calculated in the zero-current limits \cite{TH1971}, i.e. under equilibrium conditions. According to the relation $R(T)/R _{n} \approx 2.1 \alpha _{0}| \varepsilon / \varepsilon _{c}|^{9/4} \exp{ -1.89|\varepsilon / \varepsilon _{c}|^{3/2}}$ obtained in \cite{TH1971} the resistance of a homogeneous aluminium structure (with parameters of our structure $\alpha _{0} = 10.57\hbar s Gi _{1D}^{3/2}/e^{2} \xi (0) \rho _{n} \approx 1.6$) should fall down to $\approx 0.01 $ at $-\varepsilon \approx  2.6 \times \varepsilon _{c} \approx 0.0053$, down to $\approx 0.001 $ at $-\varepsilon \approx  3.2 \times \varepsilon _{c} \approx 0.0066$ and so forth. The equilibrium resistance should fall because the probability of the jumps in the normal state $P _{s \rightarrow n} \propto \exp{-\Delta F _{0}/k _{B}T} \approx \exp{-0.62|\varepsilon / \varepsilon _{c}|^{3/2}}$ diminishes with an increase of the energy difference between the normal and superconducting states  $\Delta F _{0} \approx 4\sqrt 2 \ \mu _{0} H _{c}^{2}(T) s\xi (T)/3 \approx k _{B}T 0.62|\varepsilon / \varepsilon _{c}|^{3/2}$ \cite{FlucRPP75}.

The probability $P _{s \rightarrow n} $ is negligibly small in the superconducting state at $-\varepsilon \gg \varepsilon _{c}$ under equilibrium conditions. It increases \cite{FlucRPP75} and reaches $P _{s \rightarrow n} = 1$ when a non-zero external current $I _{ext}$ exceeds the critical current $I _{c}(T) = I _{c}(0)(1 - T/T _{c})^{3/2}$. Therefore, a non-equilibrium ac current, rather than thermal fluctuations, can switch segments of a one-dimensional superconductor between the superconducting and normal states at $-\varepsilon \gg \varepsilon _{c}$ when its amplitude exceeds the critical current $I _{c}(T) $. It was discovered that such switching can induce a dc voltage $ V _{dc}$ oscillating in magnetic field $B$ in superconducting asymmetric rings \cite{Letter07,Dubonos02, PCJETP07,Letter2003} and percolating films \cite{rect1990}. The oscillations $V _{dc}(B)$ observed in asymmetric aluminium rings \cite{Letter07,Dubonos02, PCJETP07,Letter2003} are similar to the oscillations of the persistent current observed on such rings. The period of these alternating-sign oscillations $B _{0}$ corresponds to the flux quantum $B _{0}S = \Phi _{0}$ inside the ring with the area $S$ whereas the $V _{dc}$ sign \cite{Letter07,Dubonos02, PCJETP07,Letter2003} and direction of the persistent current \cite{PCScien07} change at $BS = n\Phi _{0}$ and $BS = (n+0.5)\Phi _{0}$. The observations of the oscillations $V _{dc}(BS/\Phi _{0})$ suggest that the persistent voltage $V _{p}(BS/\Phi _{0})$  can arise but they cannot be interpreted as an unambiguous experimental evidence of this equilibrium phenomena because the dc voltage $V _{dc}$ at $-\varepsilon \gg \varepsilon _{c}$ was induced by a non-equilibrium noise $I _{noise} = \int _{fmin}^{fmax} df I _{f} \sin (2\pi f + \phi _{f})$ \cite{Letter07,Dubonos02}  or sine current $I _{sine} = I _{f} \sin (2\pi f )$ \cite{PCJETP07,Letter2003}.

This non-equilibrium phenomenon may be considered as a rectification effect, i.e. the transformation of the ac current power $W _{ac} = \overline{RI^{2}}$ into the power of the dc voltage $W _{dc} = V _{dc}^{2}/\overline{R}$. The dc voltage, as well as the average value of the persistent current \cite{PCScien07}, equals zero at $\Phi = n\Phi _{0}$, $\Phi = (n+0.5)\Phi _{0}$ and has a maximum $V _{A}$ between these values of a magnetic flux, see Fig.2. The amplitude $V _{A}$  of the $V _{dc}(BS/\Phi _{0})$ oscillations does not depend on the frequency $f$ of the sine current $I _{sine} = I _{f} \sin (2\pi f )$ (at least in the frequency range $100 \ Hz - 1 \ MHz$ at $T < 0.99T _{c}$) \cite{Letter2003} and the maximum $V _{A}/R _{n}\overline{I^{2}}^{1/2}$ of the rectification efficiency $(W _{dc}/W _{ac})^{1/2} > |V _{dc}|/R _{n}\overline{I^{2}} ^{1/2}$ has the same value for both noise current with the amplitude $(\int _{fmin}^{fmax} df I _{f}^{2})^{1/2} =  \overline{2I _{noise}^{2}}^{1/2}$ and sine current with the same amplitude $\overline{2I _{sine}^{2}}^{1/2} = I _{f}$.

\begin{figure}
\includegraphics{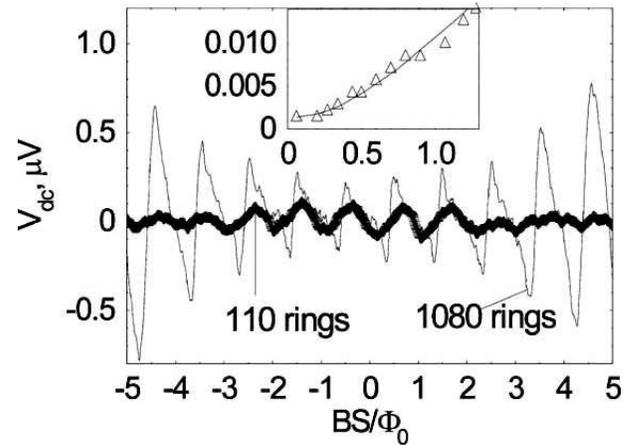}
\caption{\label{fig:epsart} Voltage oscillations in magnetic field, $V _{dc}(BS/\Phi _{0})$, measured in 110 rings at the temperature $T \approx 1.358 \ K$ corresponding to a maximum amplitude for this system with $T_{c} \approx  1.362 \ K$ and in 1080 rings at $T \approx 1.360 \ K$  below this temperature for this system with $T_{c} \approx  1.375 \ K$. The dc voltage was induced in the both cases with the same controllable ac current with the amplitude $\sqrt {2 \overline {I^{2}}} \approx 60 \ nA$. The inset shows the dependence of the maximum rectification efficiency $Eff _{Re} = V _{A,max}/R _{n} \surd 2 \overline{I^{2}}^{1/2}$ (vertical reference axis) on the amplitude $\surd 2 \overline{I^{2}}^{1/2}$ of the controllable ac current (horizontal reference axis, microampere, $\mu A$) measured on the 110 ring system.}
\end{figure}

The dc voltage $V _{dc}(BS/\Phi _{0})$ is observed at the amplitude of the ac current $\surd 2 \overline{I^{2}}^{1/2} \geq  I _{c}$ \cite{PCJETP07} when the sine or noise current can switch a ring or its segments to the normal state. The critical current $I _{c}$ of the ring equal to $I _{c}(T) = I _{c}(0)(1 - T/T _{c})^{3/2}$ at $B = 0$ changes periodically in magnetic field and decreases at high $B $, see Fig.6 and 9 of \cite{PCJETP07}. The oscillations $V _{dc}(BS/\Phi _{0})$ appear initially at a high magnetic flux inside the rings, see Fig.2 of \cite{Letter2003} and Fig.11 of \cite{PCJETP07}, because $I _{c}$ decreases with increasing $B $. The amplitude $V _{A}$ of these oscillations is a nonmonotonic function of both ac current amplitude $\surd 2 \overline{I^{2}}^{1/2}$ (at a given temperature $T$), see Fig.6 of \cite{Letter2003} and Fig.12 of \cite{PCJETP07}, and temperature (at a given $\surd 2 \overline{I^{2}}^{1/2}$ value), see Fig.6 of \cite{Letter07}. The oscillations $V _{dc}(BS/\Phi _{0})$ are not observed at low amplitudes of ac current, Fig.6 of \cite{Letter2003}, Fig.12 of \cite{PCJETP07}, and low temperatures, Fig.6 of \cite{Letter07}, because the rings cannot be switched in the normal state at $\surd 2 \overline{I^{2}}^{1/2} < I _{c}$. The amplitude $V _{A}$ decreases down to an immeasurably small value at high temperatures $T \geq T _{c}$, Fig.6 of \cite{Letter07}, because of a decrease of the persistent current \cite{PCScien07}. The maximum amplitude $ V _{A,max}$ of the oscillations $V _{dc}(BS/\Phi _{0})$ observed at $\surd 2 \overline{I^{2}}^{1/2} = I _{A,max}$ corresponds to the maximum rectification efficiency $Eff _{Re} = \surd 2 V _{A,max}/R _{n}I _{A,max}$.  

The rectification efficiency $Eff _{Re}$ allows the estimation of the minimum amplitude $\surd 2 \overline{I^{2}}^{1/2} = \surd 2 V _{A,max}/R _{n} Eff _{Re}$ of the ac current capable of inducing $V _{dc}(BS/\Phi _{0})$ oscillations with an amplitude $V _{A,max} > V _{mab}$, high enough for real observations. The measurements \cite{PCJETP07} have shown that the rectification efficiency of asymmetric aluminium rings is abnormally high in the superconducting state $Eff _{Re} = \surd 2 V _{A,max}/R _{n}I _{A,max} \approx  0.33$ at $T < 0.98T _{c}$ because of the hysteresis of the current - voltage characteristics in this temperature region. A noise with the amplitude down to $\surd 2 \overline{I^{2}}^{1/2} \approx \surd 2 V _{mab}/R _{n} Eff _{Re} \approx 0.1 \ nA = 10^{-10} \ A$ could be detected with the help of a 110 ring system ($R_{n} \approx  970 \ \Omega $)  at  the measurable voltage $V _{mab} \approx  20 \ nV $ and the above value of the rectification efficiency. But the critical current of this system $I _{c} \approx 650 \ \mu A(1 - T/T _{c})^{3/2} > 1800 \ nA$ at $T < 0.98T _{c}$ is much higher than $0.1 \ nA $. The rectification efficiency decreases at $T \rightarrow T _{c}$, Fig.15 of \cite{PCJETP07}, and thus at $\surd 2 \overline{I^{2}}^{1/2} = I _{A,max} \rightarrow 0$, Fig.2. In contrast to the critical current, the value $Eff _{Re}$ does not fall down to zero at $T \rightarrow T _{c}$. Therefore it is important that the amplitude $V _{A,max}$ of the oscillations $V _{dc}(BS/\Phi _{0})$ induced by ac current of the same amplitude $\surd 2 \overline{I^{2}}^{1/2}$ should increase with the number of rings, Fig.2. Both the amplitude $V _{A,max}$ and the resistance $R _{n}$ increase approximately proportionally to the number of rings. Therefore the systems of 110 and 1080 rings have the same maximum rectification efficiency $Eff _{Re} = V _{A,max}/R _{n}\overline{I^{2}} ^{1/2} \approx 0.002$ at the same amplitude $\surd 2 \overline{I^{2}}^{1/2} \approx  60 \ nA$. Our measurements confirmed that the system of 1080 rings can detect a noise with a lower amplitude $\surd 2 \overline{I^{2}}^{1/2} = \surd 2 V _{mab}/R _{n} Eff _{Re}$ than the system of 110 rings.

\begin{figure}
\includegraphics{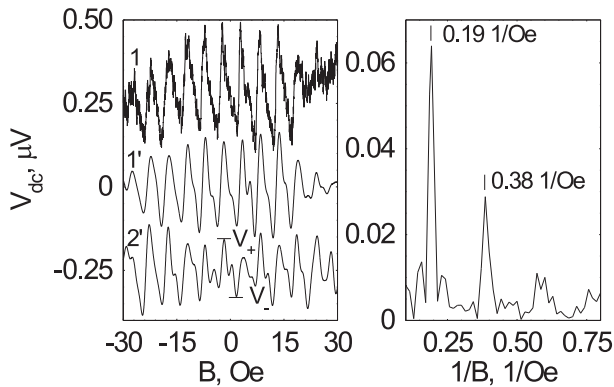}
\caption{\label{fig:epsart} Magnetic dependence of the dc voltage, $V _{dc}(BS/\Phi _{0})$, (on the left side) and the Fourier transform (on the right side). The peak of the Fourier transform at $1/B \approx 0.19 \ 1/Oe \approx 1/B _{0} \approx \pi r^{2}/\Phi _{0}$ is evidence of the quantum oscillations observed on the rings with radius $r \approx 1.1 \ \mu m$. The curve 1 (on the left side) is an original dependence measured in the 1080 rings. The oscillations 1' and 2' are the reverse Fourier transform taken in a limit region $1/B = 0.12 - 0.5 \ 1/Oe$. The temperature of 1 and 1' measurement $T \approx 1.364 \ K$ corresponds to the maximum amplitude of the quantum oscillations, Fig.4. The dependence 2' measured at $T \approx 1.362 \ K$ is typical for a lower temperature. The amplitude of the oscillations $V _{A} = (V _{+} - V _{-})/2$ was measured at low $B$. The curves 1 and 2' are vertically shifted.}
\end{figure}

The $V _{dc}(BS/\Phi _{0})$ oscillations with an amplitude to $V _{A,max} \approx 600 \ nV$ observed in \cite{Letter07} were due to an uncontrollable noise existing in any real measuring system. This non-equilibrium noise is a consequence of the galvanic connection between the parts of the measuring system located at low temperature $T \approx 1 \ K$ and at room temperature $T \approx 300 \ K$. Additional filtration with the help of low-temperature $\pi $-filters and coaxial resistive twisted pairs reduced the uncontrollable noise.  The diminution of the uncontrollable noise allowed the expansion of the calibration $Eff _{Re}(\surd 2 \overline{I^{2}}^{1/2})$ of the ring system as a noise detector \cite{Qdetector06} to the lower amplitude $\surd 2 \overline{I^{2}}^{1/2}$, Fig.2. In order to obtain the dependence $Eff _{Re}(\surd 2 \overline{I^{2}}^{1/2})$ shown in Fig.2, the maximum $V _{A,max}$ of the temperature dependence $V _{A}(T)$ was measured at each value of the amplitude $\surd 2 \overline{I^{2}}^{1/2}$ of the controllable ac current. According to this calibration the uncontrollable noise inducing the oscillations $V _{dc}(BS/\Phi _{0})$ with the maximum amplitude $V _{A,max} \approx 600 \ nV$ in \cite{Letter07} had the amplitude $\surd 2 \overline{I^{2}}^{1/2} \approx 250 \ nA$. The noise reduced with the help of  the additional filtration was undetectable with the 110 ring system but the system of 1080 rings detected it. 

To pick out barely visible quantum oscillations $V _{dc}(BS/\Phi _{0})$ in the real dependence of the voltage $V _{dc}(B)$ we used the Fourier transformation, Fig.3. This method allowed the detection of quantum oscillations with the amplitude down to $V _{mab} \approx 20 \ nV$. Our measurements made with the additional filtration showed that the amplitude $V _{A,max}$ of quantum oscillations $V _{dc}(BS/\Phi _{0})$ induced by an uncontrollable noise is $20 \ nV$ lower in the 110 ring system and reaches $ V _{A,max} \approx 100 \ nV$ in the system of 1080 rings, Fig.4. The amplitude $V _{A,max} \approx 100 \ nV $ is by $\approx 8$ time lower than the maximum amplitude $ V _{A,max} \approx 800 \ nV$ of the oscillations $V _{dc}(BS/\Phi _{0})$ observed at the controllable noise with $\surd 2 \overline{I^{2}}^{1/2} \approx 60 \ nA$. Consequently, the amplitude $ \surd 2 \overline{I^{2}}^{1/2} $ of the uncontrollable noise should be higher than $8 \ nA$ and lower than $60 \ nA$. A rough extrapolation of the dependence $Eff _{Re}(\surd 2 \overline{I^{2}}^{1/2})$, Fig.2, gives $\surd 2 \overline{I^{2}}^{1/2} \approx 20 \ nA$. Thus, low-temperature $\pi $-filters and coaxial resistive twisted pairs allowed the reduction of the amplitude of the uncontrollable noise approximately by a factor of ten.

The detection of the weak noise only with the help of the 1080 ring system demonstrates that a system of asymmetric superconducting rings is a promising detector of an uncontrollable noise. This  noise in the low-temperature part of a typical measuring system is much higher than $\surd 2 \overline{I^{2}}^{1/2} \approx 20 \ nA$. For example, the observations of $V _{dc}(BS/\Phi _{0})$ oscillations in a single aluminum ring (resistance $R _{n} \approx 15 \ \Omega $) with the amplitude $V _{A,max} \approx 1.2 \ \mu V$ \cite{Dubonos02} and in an asymmetric superconducting quantum interference device ($R _{n} \approx 2 \ \Omega $) with $ V _{A,max} \approx 15 \ \mu V$ \cite{PV1967} indicate that the amplitude of an uncontrollable noise exceeds $\surd 2 \overline{I^{2}}^{1/2} \approx 3 \ \mu A$ in the first case and $\surd 2 \overline{I^{2}}^{1/2} \approx 20 \ \mu A$ in the latter. Screening and filtration reduce the uncontrollable noise down to $\surd 2 \overline{I^{2}}^{1/2} \approx 250 \ nA$ \cite{Letter07}. Additional filtration with low-temperature $\pi $-filters and coaxial resistive twisted pairs reduce its amplitude to $\surd 2 \overline{I^{2}}^{1/2} \approx 20 \ nA$. Each time the systems with a larger number of asymmetric rings allowed the detection of a reduced non-equilibrium noise. The proportionality $V _{A,max} \propto N$ corroborated by our measurements ensures the detection of a weaker noise down to equilibrium by a system with a number of rings larger than 1080. It can be expected, for example, that $V _{dc}(BS/\Phi _{0})$ oscillations with the maximum amplitude $ V _{A,max} \approx 1 \ \mu V$, Fig.3, will be observed in the system of 10000 rings of the same type at $\surd 2 \overline{I^{2}}^{1/2} \approx 20 \ nA$ and that the noise with a lower amplitude $\surd 2 \overline{I^{2}}^{1/2} < 20 \ nA$ can be detected by this system. 

\begin{figure}
\includegraphics{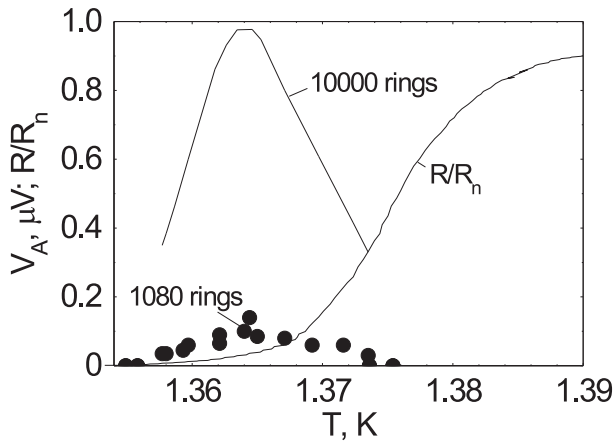}
\caption{\label{fig:epsart} Temperature dependence of the oscillation amplitude $V _{A}$ of a noise-induced voltage $V _{dc}(BS/\Phi _{0})$ measured in the 1080 rings (indicated by circles) and expected in 10000 rings. The data are shown on the background of the resistive transition ($R/R _{n}$). $R _{n} \approx 8000 \ \Omega$ is the resistance of the 1080 ring system in the normal state.}
\end{figure}

The noise current with the amplitude $\surd 2 \overline{I^{2}}^{1/2} \approx 20 \ nA$  detected with the help of 1080 rings cannot greatly exceed the Nyquist current $\surd 2 \overline{I _{Nyq}^{2}}^{1/2} \approx (k _{B}T\Delta f / R_{l,n})^{1/2}$. The frequency $f$ of the Nyquist current in the normal state is limited by the relaxation time $\tau_{RL} = L/R $ and the quantum limit $f _{QL} \approx k _{B}T /2 \pi \hbar $. The quantum limit $f _{QL} \approx k _{B}T /2 \pi \hbar \approx 2.8 \ 10^{10} \ Hz <  1/\tau_{RL} = R_{l,n}/L \approx 3 \ 10^{12} \ Hz $ is a critical factor at $T  \approx 1.36 \ K$, the ring inductance $L \approx 10^{-11} \ H$ and the resistance along the circumference $R _{l,n} \approx \rho _{n}\pi r( w _{n} + w _{w})/d w _{n}w _{w} \approx 30 \ \Omega $. The amplitude $\surd 2 \overline{I _{Nyq}^{2}}^{1/2} \approx  k _{B}T/(2 \pi \hbar R _{l,n})^{1/2} \approx 130 \ nA $ in the whole frequency band from $f = 0$ to the quantum limit $f _{QL} \approx k _{B}T /2 \pi \hbar \approx 2.8 \ 10^{10} \ Hz $ exceeds  the amplitude $\surd 2 \overline{I^{2}}^{1/2} \approx 20 \ nA$ of the noise current detected by the 1080 ring system. However there are some factors limiting the frequency of the noise current, both non-equilibrium and equilibrium, capable of inducing of both $V _{dc}(BS/\Phi _{0})$ and $V _{p}(BS/\Phi _{0})$. One of these factors is the time of relaxation to the equilibrium superconducting state. This time, equal to $\tau_{srel} \approx  \pi \hbar / 8k _{B}(T _{c} - T) $ in the linear approximation, increases near the superconducting transition, limiting the frequency band much stronger in the region where a weak noise current can induce $V _{dc}(BS/\Phi _{0})$ or $V _{p}(BS/\Phi _{0})$. Other factors can also contribute to the decrease of the frequency band and consequently the amplitude of the persistent voltage. Nevertheless, the proportionality $V _{A,max} \propto N$ should allow the detection of the persistent voltage with any amplitude when a system is made with a sufficiently large number of asymmetric superconducting rings.

We assume that the quantum oscillations $V _{dc}(BS/\Phi _{0})$ observed in the 1080 rings, Fig.3,  are  induced mainly by non-equilibrium noise because of the temperature dependence of the amplitude $V _{A}(T)$, Fig.4. The maximum of the $V _{A}(T)$ dependence is observed at $T \approx 1.364 \ K$ corresponding to the lower part of the resistive transition  $R \approx 0.03R _{n}$ whereas the maximum of the persistent voltage $V _{p}(BS/\Phi _{0})$ amplitude should be expected at $T \approx 1.376 \ K$ corresponding to the middle of the resistive transition $R \approx 0.5R _{n}$. We could observe the quantum oscillations $V _{dc}(BS/\Phi _{0})$ up to $T \approx 1.374 \ K$ where $R \approx 0.33R _{n}$. However this result does not unambiguously prove that we observed the persistent voltage in the fluctuation region because of some non-homogeneity of the system used in our work. Its resistive transition $\Delta T _{c}/ T _{c} \approx \varepsilon _{0.9} - \varepsilon _{0.1} \approx 0.015$ is wider than the one $\varepsilon _{0.9} - \varepsilon _{0.1} \approx 0.006$  expected for homogeneous system. The rings of our real system have different critical temperatures $T _{c,i}$ and different values $\varepsilon = T/T _{c,i} - 1$ at $T \approx 1.374 \ K$. Some rings with lower $T _{c,i}$ and $\varepsilon > - \varepsilon _{c} $ may contribute to the resistance $R \approx 0.33R _{n}$ whereas other rings with $\varepsilon < - 3\varepsilon _{c} $ may contribute to the quantum oscillations $V _{dc}(BS/\Phi _{0})$. The non-homogeneity of our system can partly explain the decrease of the rectification efficiency $Eff _{Re}$ with $\surd 2 \overline{I^{2}}^{1/2} \rightarrow 0$, Fig.2. A ring $i$ with its critical temperature $ T _{c,i}$ can contribute to the dc voltage $V _{dc}(BS/\Phi _{0})$ only at  $\surd 2 \overline{I^{2}}^{1/2} > I _{c}(0)(1 - T/T _{c,i})^{3/2}$ and $T < T _{c,i}$. Therefore, all rings cannot contribute to the total voltage $V _{dc}(BS/\Phi _{0})$ at any temperature $T$ when $\surd 2 \overline{I^{2}}^{1/2} < I _{c}(0)(\Delta T _{c}/T _{c})^{3/2} \approx 1 \ \mu A$ and our real system of 1080 rings with different $T _{c}$ could not detect the persistent voltage even if the maximum amplitude of $V _{p,1}(BS/\Phi _{0})$ appreciably exceeds $V _{mab}/N \approx 20 \ nV/1080 \approx 2 \ 10^{-11} \ V$.

\section{Conclusion}
In summary, we have experimentally studied the possibility to observe the persistent voltage in asymmetric superconducting rings using systems of 110 and 1080 such rings connected in series. The results of our measurements unambiguously point to the possibility that very weak electric noise can be detected with the help of such systems having a large enough number of asymmetric rings connected in series. We succeeded in reducing the non-equilibrium noise in the low-temperature part of our measuring system down to a very small power. Nevertheless, the system of 1080 rings could detect this weak noise even in spite of some imperfections of this system: its non-homogeneity, too high critical current of the rings and so on. Our results have corroborated that systems with a larger number of asymmetric rings connected in series can detect a weaker noise down to the equilibrium one. A possible observation with help of such systems the quantum oscillations of the dc voltage in the temperature region corresponding to the upper part of the resistive transition will be unambiguous evidence of the persistent voltage because non-equilibrium noise can induce a resistance but it cannot induce the persistent current. Our results suggest that the persistent voltage can be observed with the help of $\approx 10000$ rings.


\begin{thebibliography}{99}

\bibitem{PCScien09} A.C. Bleszynski-Jayich, W. E. Shanks, B. Peaudecerf, E. Ginossar, F. von Oppen, L. Glazman, J. G. E. Harris,  Science  326 (2009) 272.

\bibitem{PCPRL09} H. Bluhm, N.C. Koshnick, Ju.A. Bert, M.E. Huber, K.A. Moler, Phys. Rev. Lett. 102  (2009) 136802

\bibitem{PCScien07} N.C. Koshnick, Bluhm, H., Huber, M. E., Moler, K.A. Science 318  (2007) 1440.

\bibitem{KulikS} I.O. Kulik, Sov. Phys. JETP 31 (1970) 1172 

\bibitem{KulikN} I.O.Kulik, JETP Lett. 11 (1970) 275.

\bibitem{Riedel88} Ho-Fai Cheung, Y. Gefen, E. K. Riedel, W.H. Shih, Phys. Rev. B 37 (1988) 6050

\bibitem{Riedel92} F. von Oppen, E. K. Riedel, Phys. Rev. B 46  (1992) 3203.

\bibitem{Riedel93} E. K. Riedel, F. von Oppen, Phys. Rev. B 47 (1993) 15449 

\bibitem{Kulik75} V. L. Gurtovoi, A. I. Ilin, A.V. Nikulov, V. A. Tulin, Low Temp. Phys. 36 (2010) 974

\bibitem{Birge09} N. O. Birge, Science 326 (2009) 244.

\bibitem{FeynmanLec} R.P Feynman, R. B Leighton, M. Sands, The Feynman Lectures on Physics, Addison - Wesley Publishing, Massachusetts, 1963.

\bibitem{Nyquist} H. Nyquist, Phys. Rev. 32 (1928) 110.

\bibitem{Johnson}  J. B. Johnson, Phys. Rev. 32 (1928) 97. 

\bibitem{Letter07} A.A. Burlakov, V.L. Gurtovoi, S.V. Dubonos, A.V. Nikulov, V.A. Tulin,  JETP Lett. 86  (2007) 517; arXiv: 0805.1223 (2008). 

\bibitem{LP1962} W. A. Little, R. D. Parks, Phys. Rev. Lett. 9 (1962) 9.

\bibitem{Dubonos02} S.V. Dubonos, V.I. Kuznetsov, A.V. Nikulov, in Proceedings of 10th International Symposium "NANOSTRUCTURES: Physics and Technology", St Petersburg: Ioffe Institute, (2002) p.350; arXiv: 1112.6157 (2011).

\bibitem{JLTP1998} A.V.Nikulov, I.N.Zhilyaev, J. Low Temp. Phys. 112 (1998) 227.

\bibitem{PRB2001} A.V. Nikulov, Phys. Rev. B 64 (2001) 012505.

\bibitem{Levy1990} L. P. Levy, G. Dolan, J. Dunsmuir, H. Bouchiat, Phys. Rev. Lett. 64  (1990) 2074

\bibitem{PCJETP07} V.L. Gurtovoi, S.V. Dubonos, A.V. Nikulov, N.N. Osipov, V.A. Tulin,  Zh. Eksp. Teor. Fiz. 132 (2007) 1320; arXiv: 0903.3539 (2009)

\bibitem{FlucRPP75} W.J. Skocpol, M. Tinkham, Rep. Prog. Phys. 38 (1975) 1049 

\bibitem{FlucLL1972} R. J. Londergan, J. S. Langer, Phys. Rev. B 5 (1972) 4376

\bibitem{TH1971} J.R. Tucker, B.I. Halperin, Phys. Rev. B 3 (1971) 3768

\bibitem{Letter2003} S. V. Dubonos, V. I. Kuznetsov, I. N. Zhilyaev, A. V. Nikulov, A.A. Firsov, JETP Lett.  77 (2003) 371; arXiv: cond-mat/0303538 (2003).

\bibitem{rect1990} A. Gerber, G. Deutscher, Phys. Rev. Lett. 64  (1990) 1585

\bibitem{Qdetector06} V.L. Gurtovoi, S.V. Dubonos, A.V. Nikulov, N.N. Osipov, V.A. Tulin, in the Proceedings of  SPIE Vol. 6260, Micro- and nanoelectronics - 2005,  edited by Kamil A. Valiev and Alexander A. Oplikovskii, (2006) p. 62600T1; arXiv: cond-mat/0603009

\bibitem{PV1967} A.Th.A.M. De Waele, W.H.Kraan, R. De Bruynouboter, K.W. Taconis, Physica 37 (1967) 114.

\end{thebibliography}
\end{document}